\documentclass[aps,pra,twocolumn,nofootinbib]{revtex4-2}
\usepackage[colorlinks]{hyperref}
\usepackage[utf8]{inputenc}
\usepackage{graphicx}
\usepackage{multirow}
\usepackage{amssymb}
\usepackage{amsmath}
\usepackage{gensymb}

\begin{document}

\title{
Thermal and annihilation radiation in the quark nugget model of dark matter
}

\author{V.~V.~Flambaum$^{1,2,3}$}
\author{I.~B.~Samsonov$^1$}
\affiliation{$^1$School of Physics, University of New South Wales, Sydney 2052, Australia}
\affiliation{$^2$Helmholtz Institute Mainz, GSI Helmholtzzentrum für Schwerionenforschung, 55099 Mainz, Germany}
\affiliation{$^3$Johannes Gutenberg University Mainz, 55099 Mainz, Germany}

\begin{abstract}
The Quark Nugget (QN) model of dark matter suggests that the dark matter may consist of compact composite objects of quark matter. Although such composite particles can strongly interact with visible matter, they may remain undetected because of a small cross section to mass ratio. We focus on anti-QNs made of antiquarks since they are heated by annihilation with visible matter and radiate. We study the radiation spectrum and power from anti-QNs in our galaxy and compare them with satellite observations. Thermal radiation from  anti-QN is produced by fluctuations of the positron density. We calculate the thermal radiation  of anti-QNs with the use of the Mie theory and found its ratio to the black-body radiation. This allows us to find the equilibrium temperature of anti-QNs in the interstellar medium and determine their contribution to the observed diffuse background radiation in our galaxy in different frequency intervals, from radio to UV. We also consider non-thermal radiations from anti-QNs which are produced by products of annihilations of particles of the interstellar gas with anti-QNs. Such radiations include photons from decays of $\pi^0$ mesons, synchrotron, bremsstrahlung and transition radiations from $\pi^\pm$ mesons, electrons and positrons. Synchrotron radiation in MHz frequency range  and flux of photons from $\pi^0$ decays  may be above the detection threshold in such detectors as Fermi-LAT.
\end{abstract}

\maketitle

\section{Introduction}

The quark nugget model of dark matter suggests that the dark matter particles may be represented by compact composite objects composed of a large number of quarks or antiquarks. Such objects can strongly interact with visible matter, being cosmologically and astrophysically ``dark'' due to a small cross section to mass ratio, $\sigma/M\ll1\mbox{ cm}^2/\rm g$. Different models of this class assume different mechanisms of stabilization of the quark matter. In particular, the model of strangelets \cite{Witten84} suggests that the quark matter enriched with the strange quarks can have a lower ground state energy than the baryonic matter. Another interesting model is the Axion-Quark Nugget (AQN) model \cite{Zhitnitsky2002} which is based on the assumption that axion domain walls can create a high pressure inside small bubbles of matter which would render quarks to turn to the color-superconducting state with a lower ground state energy as compared with the baryonic state, see, e.g., \cite{ColorSuperconductivityReview}. In this paper, we study thermal radiation in such models in general, referring to them as to quark nuggets (QN), without specifying any particular model of this kind.

An interesting feature of the quark nugget model of dark matter is that dark matter particles may be represented by both quark and antiquark nuggets \cite{Zhitnitsky2002}. Assuming that the antiquark nuggets are approximately 1.5 times more abundant in the universe, it is possible to show that the dark matter to visible matter density ratio is close to the observed one 5:1 \cite{Zhitnitsky2006}. In this model, the antiquark nuggets play special role because they can manifest themselves through annihilation processes with the visible matter. Therefore, we will focus mainly on the properties of antiquark nuggets and their possible detection techniques. 

The thermal properties of antiquark nuggets are specified by the positron cloud which compensates the electric charge of the quark core. Following the ideas of the earlier work \cite{Electrosphere2008}, in Ref.~\cite{FS21} we studied the density distribution in the positron cloud within the Thomas-Fermi model. We showed that inside the quark core the positron density is very high, reaching the values $1.7\times 10^{35}\mbox{ cm}^{-3}$, while near the boundary it drops nearly by 10 orders at the distance of one $a_B$. Thus, the positron cloud may be roughly considered as a relativistic gas inside a sphere of radius $R$. Density fluctuations in this gas would produce thermal radiation from the quark nugget.

We should note that radiation of individual electrons or positrons calculated using two-body bremsstrahlung formulas as in Ref.~\cite{WMAPhaze} does not give accurate result in a dense electron or positron gas. When the radiation wavelength is bigger than the distance between particles, thermal radiation appears as a collective effect originating from fluctuations of charge density\footnote{A well known example of collective effects is plasma oscillations.}. This is well known for electron gas in metals and metal nanoparticles, see, e.g., \cite{BHbook}.

In this paper, we study the thermal radiation spectrum and power of (anti)quark nuggets by modelling them as small spherical particles with a given radius $R$ and plasma frequency $\omega_p$. In this model, the radiation power may be found with the use of Mie theory, see, e.g., \cite{BHbook}. Within this theory, the radiation power at any given temperature $T$ may be represented as $P(\omega,T)=\pi E(\omega)I_0(\omega,T)$, where $I_0(\omega,T)$ is the Planck function and $E(\omega)$ is the radiation emissivity function. We calculate this function numerically (and analytically in the low and high frequency limits) for the positron gas in the quark nugget model. This allows us to compare the quark nugget thermal radiation with the black-body radiation. 

We show that at temperatures $k_B T\sim 1$ eV thermal radiation power from (anti)quark nuggets appears nearly 100 times higher than that estimated in Ref.~\cite{WMAPhaze}. This is the result of the collective effects which have not been taken into account in the earlier calculation. The dependence of the radiation power $P(\omega,T)$ on the frequency $\omega$ is also very strongly affected by the collective effects. In Ref.~\cite{WMAPhaze} the spectrum is nearly ``flat'' in a range of frequencies while $P(\omega,T)=\pi E(\omega)I_0(\omega,T)$ has a steep decrease towards small frequencies as a combined effect of the decrease of the emissivity $E(\omega)$ and Plank  function $I_0(\omega,T)$. This is important for the explanation of such low-frequency phenomena as ``WMAP haze''  \cite{WMAP1,WMAP2,WMAP3,WMAP4}. Based on this result, we revisit the radiation pattern of quark nuggets in our galaxy and compare it with satellite observations in different spectral bands.

Apart from the thermal radiation, antiquark nuggets may produce other types of radiation which originate from matter-antimatter annihilations when QNs collide with particles of interstellar gas. In Ref.~\cite{FS21} we considered emission of 511 keV photons from the electron-positron annihilation. However, the annihilation processes  yield also the emission of $\pi$ mesons (as well as $\gamma$ rays) which further decay into energetic photons, neutrinos, electrons and positrons. Thus, QNs produce cosmic rays which may be, potentially, detected. In this paper, we consider the synchrotron radiation from these electrons and positrons in our galaxy, as well as bremsstrahlung and transition radiation from charged $\pi$ mesons.

The rest of the paper is organized as follows. In Sect.~\ref{Parameters} we review the basic aspects of the quark nugget model which are employed in the subsequent sections. In Sect.~\ref{sec-radiation}, we calculate the thermal emissivity of (anti)quark nuggets and compare their radiation power with the black-body radiation. In Sect.~\ref{Equilibrium} we determine the thermal equilibrium temperature for QNs in different media. This allows us to study thermal and annihilation radiation from QNs in the interstellar medium in our galaxy and compare it with different satellite observations. This is the subject of Sect.~\ref{Galaxy}. Finally, in Sect.~\ref{Summary}, we give a summary of the results of this work.

Throughout this paper we use natural units in which $\hbar=1$ and $c=1$.

\section{Parameters of the quark nugget model}
\label{Parameters}

Quark nuggets are characterized by the following macroscopic parameters: size, mass, electric charge and baryon charge number. All these parameters are not known precisely, but there are preferred values of these parameters dictated by the known properties of the dark matter. In this section, we briefly discuss these values in the parameter space referring mainly to the works \cite{Zhitnitsky2002,Survival,IceCube,ColorSuperconductivityReview,CFL-charge,Electrosphere2008,FS21}.  Although this section contains no new results, we keep it for consistency of the paper.

The main assumption in the QN model is that the quark nuggets are compact composite objects composed of a large number of quarks. Within the axion quark nugget (AQN) framwork \cite{Zhitnitsky2002} it is additionally assumed that the stability of such objects is ensured by a color superconducting state of quarks with lower energy and higher density as compared with the baryonic state. This means that the matter density in quark nuggets is on the order of the nuclear matter, and can slightly exceed it. This statement amounts to the following relation between the characteristic radius and baryon charge number of QNs:
\begin{equation}
    R\simeq B^{1/3} \times 1\mbox{ fm}\,.
    \label{radius-relation}
\end{equation}

Quark nuggets may start their formation in the early universe, when the temperature was on the order of the QCD scale, $T\sim \Lambda_{\rm QCD}$. Since they strongly interact with matter, they could fully annihilate in the dense primordial plasma if they carried a small baryon charge number. In Ref.~\cite{Survival}, it was shown that quark nuggets could survive until present time if they possess the baryon charge number 
\begin{equation}
    B \gtrsim 10^{24}\,.
    \label{Bconstraint}
\end{equation}
A stronger constraint $B > 3\times 10^{24}$ was found in Ref.~\cite{Survival} based on the non-observation of any manifestations of quark nuggets in the IceCube experiment \cite{IceCube}. When combined with the relation (\ref{radius-relation}), the condition (\ref{Bconstraint}) implies $R>1.4 \times 10^{-7}$ m. In this paper, we will be a little more conservative assuming that the characteristic radius of quark nuggets is 
\begin{equation}
    R= 10^{-7}\mbox{ m},
    \label{radius}
\end{equation} 
that corresponds to $B=10^{24}$. In reality, quark nuggets, if exist, would have a size distribution with the central value (\ref{radius}). Throughout this paper, for rough estimates we will assume that all QNs have the same size (\ref{radius}).

The baryon charge of quark nuggets has an upper bound \cite{BFZ}
\begin{equation}
    B \lesssim 10^{28}\,.
    \label{UpperConstraint}
\end{equation}
This constraint was obtained as an re-interpretation of the results of the work \cite{Herrin2005} that at least 90\% of quark nuggets should have mass below 10 kg. In this paper, we will focus mainly on the window $10^{24}\lesssim B \lesssim 10^{28}$ and  will use the lower value $B=10^{24}$ for our estimates, although, as we will show below, the sensitivity to this parameter is weak.

Another important parameter in the QN model is the electric charge of the quark core. Unfortunately, the value of this parameter is model dependent and may significantly vary in different scenarios. In Ref.~\cite{Zhitnitsky2002} it was argued that at high pressure and low temperature the quarks form a condensate of Cooper pairs known as the colour superconducting state. However, there may be many different phases of color superconductors, depending on the quark pairing pattern, see, e.g.~\cite{ColorSuperconductivityReview} for a review. In particular, in the so-called color-flavor locked (CFL) phase, the three quarks (up, down and strange) are present in nearly equal numbers, and the electric charge number $Z$ arises as a surface effect, $Z\approx 0.3 B^{2/3}$ \cite{CFL-charge}. Another important phase is the so-called two flavor color superconducting (2CS) phase, in which only two lightest quarks (up and down) play role. In this phase, the electric charge is non-trivial in the bulk of the quark core, and its value may be found from the requirement of beta equilibrium, $Z\approx 5\times 10^{-3}B$, see \cite{ColorSuperconductivityReview} for details. Although both these phases are of interest, in this paper we will consider only the 2CS phase, leaving other cases for separate studies. 

In vacuum, quark nuggets must be electrically neutral. Therefore, the quark core must be surrounded by the electron cloud in the case of quark nuggets, or by the positron cloud in the case of anti-quark nuggets. The distribution of the electric charge in the electron (or positron) cloud may be found by solving the Thomas-Fermi equation. In Ref.~\cite{Electrosphere2008} this problem was studied for the QNs with the CFL core, while similar solutions for the 2CS core were found in \cite{FS21}. We will focus on the latter case and consider mainly antiquark nuggets where the positron density is usually denoted by $n_{e^+}$.

Within the Thomas-Fermi model, the positron (or electron) density is related to the chemical potential $\mu$ as
\begin{equation}
\label{ne-density}
    n_{e^+} = \frac1{3\pi^2}[(\mu+m)^2-m^2]^{3/2}\,,
\end{equation}
with $m$ being the electron mass. The value of the chemical potential (which is identified here with the Fermi energy, $\mu\equiv\varepsilon_F$) inside quark nuggets is dictated by the beta equilibrium. This value may vary in the wide range from 10 to 100 MeV, depending on the phase of the quark core \cite{StrangeStars}. In Ref.~\cite{FS21} the Thomas-Fermi equation was solved with different boundary conditions for the chemical potential in this interval, and it was shown that macroscopic properties of the positron do not change significantly for these values. Therefore, in this paper we will accept an intermediate value $\mu_R = 25$ MeV on the boundary of QN, which corresponds to the chemical potential 
\begin{equation}
    \mu = 33.5 \mbox{ MeV}
\end{equation}
inside the quark core (see Table I in Ref.~\cite{FS21}). This value of the chemical potential corresponds to the total charge number of the positron cloud $Z = 7.2\times 10^{20}$. 

We point out that the values of the macroscopic parameters of quark nuggets discussed here are known only approximately. These values are suitable for order-of-magnitude estimates that is sufficient for the goals of this work.


\section{Thermal radiation from (anti)quark nuggets}
\label{sec-radiation}

\subsection{Thermal radiation from small particles}

The spectral density of the black body radiation is known to be described by the Planck function
\begin{equation}
    I_0(\omega,T) = \frac{\hbar\omega^3}{4\pi^3 c^2}
    \frac1{\exp(\hbar\omega/(k_B T))-1}\,,
    \label{I0}
\end{equation}
with $k_B$ being the Boltzmann constant. In what follows, we include the Boltzmann constant into the definition of temperature, $k_B T \to T$, meaning that the temperature is measured in units of energy (eV). The function (\ref{I0}) quantifies the flux of energy per unit time from the black body at temperature $T$ in the frequency interval $(\omega,\omega+d\omega)$ through the surface element $ds$ normal to the direction of propagation.

To find the radiation power from infinitesimal surface element $ds$ per unit frequency interval $\omega$ in all directions, one integrates the function (\ref{I0}) over the angles with the weight $\cos\theta$,
\begin{equation}
\begin{split}
    P_0(\omega,T)& = \int_0^{2\pi}d\varphi \int_0^{\pi/2}
    I_0(\omega,T)\cos\theta\sin\theta d\theta \\
    &=\pi I_0(\omega,T)\,.
    \label{P0}
\end{split}
\end{equation}
The total radiation power per unit surface area is
\begin{equation}
    W_0(T) = \int_0^\infty P_0(\omega,T) d\omega = \sigma T^4\,,
    \label{W0}
\end{equation}
where $\sigma$ is the Stefan-Boltzmann constant.

According to the Kirchhoff's law of thermal radiation, for arbitrary body the radiation power per unit area at any given frequency $P(\omega,T)$ is proportional to the black body radiation power (\ref{P0}),
\begin{equation}
    P(\omega,T) = E(\omega) P_0(\omega,T)\,,
    \label{P}
\end{equation}
where $E(\omega)$ is the emissivity coefficient. As is shown in Ref.~\cite{BHbook}, the thermal emissivity of a small spherical particle of radius $R$ is expressed via the cross sections of extinction and scattering (reflection) of radiation,
\begin{equation}
    E(\omega) = \frac1{\pi R^2} (\sigma_{\rm ext}(\omega) - \sigma_{\rm scatt}(\omega))\,.
    \label{E}
\end{equation}
These cross sections are represented by the following series expansions
\begin{subequations}
\label{calc-details}
\begin{eqnarray}
\sigma_{\rm scatt}(\omega) &=& \frac{2\pi}{(\omega/c)^2} 
 \sum_{n=1}^\infty (2n+1) (|a_n(\omega)|^2 + |b_n(\omega)|^2)\,,~~~~~\\
\sigma_{\rm ext}(\omega) &=& \frac{2\pi}{(\omega/c)^2} 
 \sum_{n=1}^\infty (2n+1) {\rm Re}[a_n(\omega) + b_n(\omega)]\,,
\end{eqnarray}
\end{subequations}
where
\begin{subequations}
\label{coefficients}
\begin{eqnarray}
a_n(\omega) &=& \frac{N \psi_n(Nx)\psi'_n(x)-\psi_n(x)\psi'_n(N x)}{N\psi_n(Nx)\xi'_n(x) - \xi_n(x)\psi'_n(Nx)} \,,\\
b_n(\omega) &=& \frac{ \psi_n(Nx)\psi'_n(x) - N\psi_n(x)\psi'_n(N x)}{\psi_n(Nx)\xi'_n(x) - N\xi_n(x)\psi'_n(Nx)}
\,.
\end{eqnarray}
\end{subequations}
Here
\begin{equation}
    \psi_n(x) = x j_n(x)\,,\quad
    \xi_n(x) = x h^{(1)}_n(x)
    \label{RiccatiBessel}
\end{equation}
are the Riccati-Bessel functions, $N$ is the relative refractive index, and $x = \omega R/c$. 

The total radiation power per unit surface area of QN is given by the equation similar to (\ref{W0}),
\begin{equation}
    W(T) = \int_0^\infty P(\omega,T) d\omega = \int_0^\infty E(\omega)P_0(\omega,T) d\omega\,.
    \label{W}
\end{equation}
To facilitate the comparison with the black body radiation power (\ref{W0}), Eq.~(\ref{W}) may be represented in the form
\begin{equation}
    W(T) = R(T) \sigma T^4\,,
    \label{WR}
\end{equation}
where the function
\begin{equation}
    R(T) = \frac1{\sigma T^4} \int_0^\infty E(\omega)P_0(\omega,T) d\omega
    \label{R}
\end{equation}
describes the deviation from the black body radiation power.

In a non-magnetic medium, the relative refractive index is expressed via the relative permittivity (dielectric constant) $\varepsilon(\omega)$,
\begin{equation}
    N(\omega) = \sqrt{\varepsilon(\omega)/\varepsilon_0}\,,
    \label{N}
\end{equation}
where $\varepsilon_0=1$ is the vacuum permettivity. The relative permittivity is the main parameter specifying the radiation from the particle. In the next section, we will find this function in the quark nugget model.

\subsection{Dielectric constant in the quark nugget model}
\label{DielectricConstant}

In the QN model, the electron (or positron) cloud is represented by the degenerate relativistic Fermi gas with typical Fermi momentum $p_F=34$ MeV. At non-zero temperature, the electrons may occupy excited states above the Fermi level. These electrons interact with the electromagnetic radiation and, thus, are responsible for the electric conductivity. Therefore, it is appropriate to study the dielectric properties of QNs by analogy with metals. In particular, a satisfactory description for the dielectric constant may be achieved within the Drude model (see, e.g., \cite{BHbook}),
\begin{equation}
    \varepsilon(\omega) = 1-\frac{\omega_p^2}{\omega^2 + i\gamma \omega}\,,
    \label{epsilon}
\end{equation}
where $\omega_p$ is the plasma frequency and $\gamma$ is the damping constant. In this section we will estimate these parameters for the electron cloud in the quark nugget model.

\subsubsection{Plasma frequency}
In a degenerate relativistic Fermi gas, the plasma frequency squared is defined by (see, e.g., \cite{Delsante})
\begin{equation}
    \omega_p^2 = \frac{4\pi\alpha}{3} \frac{\partial n_e}{\partial\mu}\,, 
\end{equation}
where $\alpha$ is the fine structure constant, $n_e$ is the electron (or positron) density (\ref{ne-density}), $m$ is the electron mass and $\mu$ is the chemical potential in the Thomas-Fermi model. Thus, in the ultrarelativistic case ($\mu\gg m$), it reads
\begin{equation}
    \omega_p\approx 2\sqrt{\frac{\alpha}{3\pi}}\mu\,.
    \label{omega-p}
\end{equation}

The electron (or positron) gas inside the quark nugget is subject to very high density. The typical value of the chemical potential for this gas was estimated in Ref.~\cite{FS21}, $\mu=33.5$ MeV. With this value of the chemical potential, we estimate the plasma frequency in the quark nugget model,
\begin{equation}
    \omega_p\approx 2 \mbox{ MeV}.
    \label{omegap}
\end{equation}
This value is very large as compared with the one in metals, where it is of order of ten eV.

\subsubsection{Damping constant}

A proper calculation of the damping constant $\gamma$ is a complicated many-body problem which should be considered in future. However, as we will see below, dependence of the QN radiation on the damping constant $\gamma$ is very weak. For example, temperature of QN in equilibrium (radiated  energy equal to cosmic hydrogen annihilation energy on QN)  is $T \propto \gamma^{-1/9}$. Therefore, in this paper we only perform a simple single-particle estimate of $\gamma$.

In the Drude model, the damping constant $\gamma$ is identified with the collision frequency of electrons. In metals, main contributions to $\gamma$ come from the electron scattering off the lattice phonon oscillations and due to defects and impurities. Thus, to estimate $\gamma$ in the QN model we have to specify the properties of quark core and its structure.

In Ref.~\cite{Zhitnitsky2002} it was conjectured that the quarks in the quark core condense into a color-superconducting state at very high pressure and low temperature. Such state may have an energy gap which would prevent the low-energy core excitations to occur. In this paper, we consider a more general model, in which we do not specify the state of the quark core. To estimate the damping constant, we just assume that the electrons (or positrons) can scatter off particular quarks or (anti)protons with a charge density $n_{\rm core}$. The latter may be estimated simply as $n_{\rm core} = Z_{\rm core}/(\frac{4\pi}3 R^3)$, with $Z_{\rm core}$ being the total charge number of the core. This number was estimated in Ref.~\cite{FS21}, $Z_{\rm core}=7.2\times 10^{20}$ for the total baryon number $B=10^{24}$. Representing the QN radius as $R=1{\rm fm}\times B^{1/3}$, we find
\begin{equation}
    n_{\rm core} = 1.7\times 10^{-4} {\rm fm}^{-3}\,.
\end{equation}

For ultrarelativistic electrons (or positrons), the damping constant may be expressed via the scattering cross section,
\begin{equation}
    \gamma = n_{\rm core}\sigma\,.
    \label{gamma-sigma}
\end{equation}
 To estimate $\sigma$, we assume that the electron (or positron) scatters off a heavy (anti)proton or quark with the screened Coulomb (Yukawa) potential. The inverse Debye screening length was estimated in Refs.~\cite{FS21,Electrosphere2008},
\begin{equation}
    m_{\rm eff} \equiv \lambda_D^{-1} = \frac{2e}{\sqrt{\pi}} \mu = 3.2\mbox{ MeV}\,,
\end{equation}
where we used the value of the chemical potential inside the quark nugget $\mu = 33.5$ MeV. With this value of the screening lengths, we can employ the differential cross section of the relativistic electron scattering off the Yukawa potential \cite{LL4},
\begin{equation}
    \frac{d\sigma}{d\Omega} = \frac{4\alpha^2 (p^2+m^2)}{(m_{\rm eff}^2 + 4p^2 \sin^2\frac{\theta}{2})^2} \left(1 - \frac{p^2 \sin^2\frac\theta2}{p^2+m^2}
    \right).
\end{equation}
In the ultrarelativistic case, $p\gg m$, this formula gives the full cross section in the simple form
\begin{equation}
    \sigma(p) = \pi \alpha^2 \left[
     \frac4{m_{\rm eff}^2} - \frac1{p^2} \ln\left(1+\frac{4p^2}{m_{\rm eff}^2}\right)
    \right].
    \label{electron-scattering}
\end{equation}

Since only the electrons (positrons) in the vicinity of the Fermi surface can scatter, we have to consider the cross section (\ref{electron-scattering}) at $p=p_F = 34$ MeV,
\begin{equation}
    \sigma(p_F) = 0.014\ {\rm fm}^2\,.
\end{equation}
Substituting this value of the cross section into Eq.~(\ref{gamma-sigma}) we get the estimate for the damping constant:
\begin{equation}
    \gamma = 475\mbox{ eV}.
    \label{gamma}
\end{equation}

This simple estimate for the damping constant is sufficient to achieve the goals of this paper in obtaining conservative results regarding the radiation power of quark nuggets. More generally, one could take into account other contributions to $\gamma$, such as the interaction of positrons  with positron  density waves (Landau damping). 


\subsection{Thermal emissivity of quark nuggets}
\label{sec-emissivity}

Having found the plasma frequency (\ref{omegap}) and the damping constant (\ref{gamma}), we are prepared to study the thermal emissivity of quark nuggets using Eq.~(\ref{E}). In general, the series (\ref{calc-details}) can be calculated numerically for any given value of the frequency $\omega$. However, these expressions may be significantly simplified for long and short wavelengths. 

First, we note that  the values of plasma frequency (\ref{omegap}) and damping constant (\ref{gamma}) are unnaturally high as compared with the ones for ordinary metals and dielectrics. As a result, the refractive index (\ref{N}) calculated with these parameters is very large, $|N(\omega)|\gg1$, in a very broad frequency interval $\omega < \omega_p/10$.
Therefore, we can use the asymptotic forms of the Riccati-Bessel functions (\ref{RiccatiBessel}) 
$\psi_n(z)\approx \sin(z-\pi n/2)$ with $ |z|\gg n$ in order to simplify the expressions for the coefficients (\ref{coefficients}):
\begin{subequations}
\label{coefficients-approx}
\begin{eqnarray}
a_n(\omega) &\approx& \frac{ iN(\omega) \psi'_n(\omega R/c)-\psi_n(\omega R/c)}{iN(\omega)\xi'_n(\omega R/c) - \xi_n(\omega R/c)} \,,\\
b_n(\omega) &\approx& \frac{ i\psi'_n(\omega R/c) - N(\omega)\psi_n(\omega R/c)}{i\xi'_n(\omega R/c) - N(\omega)\xi_n(\omega R/c)}
\,.
\end{eqnarray}
\end{subequations}
With these coefficients, the QN emissivity (\ref{E}) may be written as 
\begin{equation}
\label{E-simpl}
\begin{aligned}
    E(\omega) &= \frac{2c^2}{(\omega R)^2} \sum_{n=1}^\infty (2n+1)
    \Big\{ 
    {\rm Re}[ a_n(\omega) + b_n(\omega)] 
    \\
&   \qquad  - |a_n(\omega)|^2
    -|b_n(\omega)|^2
    \Big\}\,.
\end{aligned}
\end{equation}
This expression for the emissivity is suitable for numerical calculations. However, further simplifications may be achieved for long and short wavelengths.

\subsubsection{Short wavelength limit}

For short wavelengths with $\omega R/c > n$, we employ the asymptotic behavior of the Riccati-Bessel functions $\psi_n(\omega R/c)\approx \sin(\omega R/c-\pi n/2)$, $\xi_n(\omega R/c) \approx \exp[i(\omega R/c - \frac{n+1}{2}\pi)]$, to further simplify the coefficients (\ref{coefficients-approx}). As a result, it may be shown that 
\begin{equation}
\label{shortwaves}
{\rm Re}[a_n(\omega)+b_n(\omega)]-|a_n(\omega)|^2-|b_n(\omega)|^2
\approx 2{\rm Re}[\zeta(\omega)]\,,
\end{equation}
where 
\begin{equation}
    \zeta(\omega) = \sqrt{\frac{\varepsilon_0 }{\varepsilon(\omega)}}
\end{equation}
is the impedance and $\varepsilon_0=1$ is the vacuum permittivity.

For $\omega R/c <n$, however, the coefficients (\ref{coefficients-approx}) have large denominators due to $|\xi_n(R\omega/c)|\ll1$ and $|\xi'_n(R\omega/c)|\ll1$. Therefore, the sum in Eqs.~(\ref{calc-details}) may be effectively cut off at $n=\omega R/c$, $\sum_{n=1}^{\omega R/c} 1= (\omega R/c)^2 + \omega R/c +1 \approx (\omega R/c)^2$. As a result, making use of Eq.~(\ref{shortwaves}), we find the following asymptotic behaviour of the emissivity:
\begin{equation}
    E(\omega) =     5.36 \, {\rm Re}[\zeta(\omega)] \,.
    \label{emissivity-asympt}
\end{equation}
The coefficient 5.36 was calculated numerically by matching the asymptotic function (\ref{emissivity-asympt}) with the exact result (\ref{E})--(\ref{coefficients}) in a range of frequencies $\omega \gtrsim 50\mbox{ eV} $.

Practically, for QNs with size parameter $R=10^{-7}$ m, Eq.~(\ref{emissivity-asympt}) describes the emissivity with good accuracy for $20\mbox{ eV}<\omega < 2$ MeV.

\subsubsection{Long wavelength limit}
In the long wavelength limit, $\omega R/c\ll1$, one can use series decomposition of the Riccati-Bessel functions (\ref{RiccatiBessel}) to show that the leading terms in the coefficients (\ref{coefficients-approx}) are given by
\begin{subequations}
\begin{eqnarray}
    {\rm Re}[a_n(\omega)] - |a_n(\omega)|^2 &\approx& (\omega R/c)^4 {\rm Re}[\zeta(\omega)]\,,\\
    {\rm Re}[b_n(\omega)] - |b_n(\omega)|^2 &\approx& (\omega R/c)^2 {\rm Re}[\zeta(\omega)]\,.
\end{eqnarray}
\end{subequations}
As a result, the emissivity (\ref{E}) may be approximated as
\begin{equation}
    E(\omega) \approx 6 (1+ (\omega R/c)^2){\rm Re}[\zeta(\omega)]\,.
    \label{ELow}
\end{equation}
For quark nuggets with characteristic size $R=10^{-7}$ m, Eq.~(\ref{ELow}) is applicable for frequencies $\omega<1$ eV.

For low frequencies, $(\omega R/c)^2\ll 1$, Eq.~(\ref{ELow}) may be further simplified and written explicitly in terms of the plasma frequency and the damping constant,
\begin{equation}
    E(\omega) \approx  6 \,  {\rm Re}[\zeta(\omega)] = 6 {\rm Re} \left(\frac{\omega^2 + i \gamma \omega}{\omega^2 -\omega_p^2 + i \gamma \omega}\right)^{1/2}.
    \label{Egeneral}
\end{equation}
It is interesting to note that this formula is very close to the short wavelength approximation (\ref{emissivity-asympt}). Thus, with $\sim$ 10\% accuracy it may be used as an interpolation formula for all frequencies $\omega \ll\omega_p$. Furthermore, in the range of frequencies $\omega \ll \gamma \ll \omega_p$ we have ${\rm Re}[\zeta(\omega)] \approx \sqrt{\omega \gamma/(2 \omega_p^2})$, and Eq.~(\ref{Egeneral}) may be cast in the form
\begin{equation}
    E(\omega) \approx   3  \frac{\sqrt{2\omega \gamma}}{\omega_p}\,.
    \label{Eapprox}
\end{equation}
This simple expression for thermal emissivity allows us to obtain an explicit expression for the ratio of the QN radiation power to the black body radiation power (\ref{R})
\begin{equation}
    R(T) = \frac{W(T)}{W_0(T)} \approx 8.0\frac{\sqrt{T \gamma}}{\omega_p}\,.
    \label{RT}
\end{equation}
This relation is applicable only for low QN effective temperature $T\ll \gamma$.

\subsubsection{Discussion of the emissivity for QNs}

\begin{figure*}[htb]
    \centering
    \includegraphics[width=10cm]{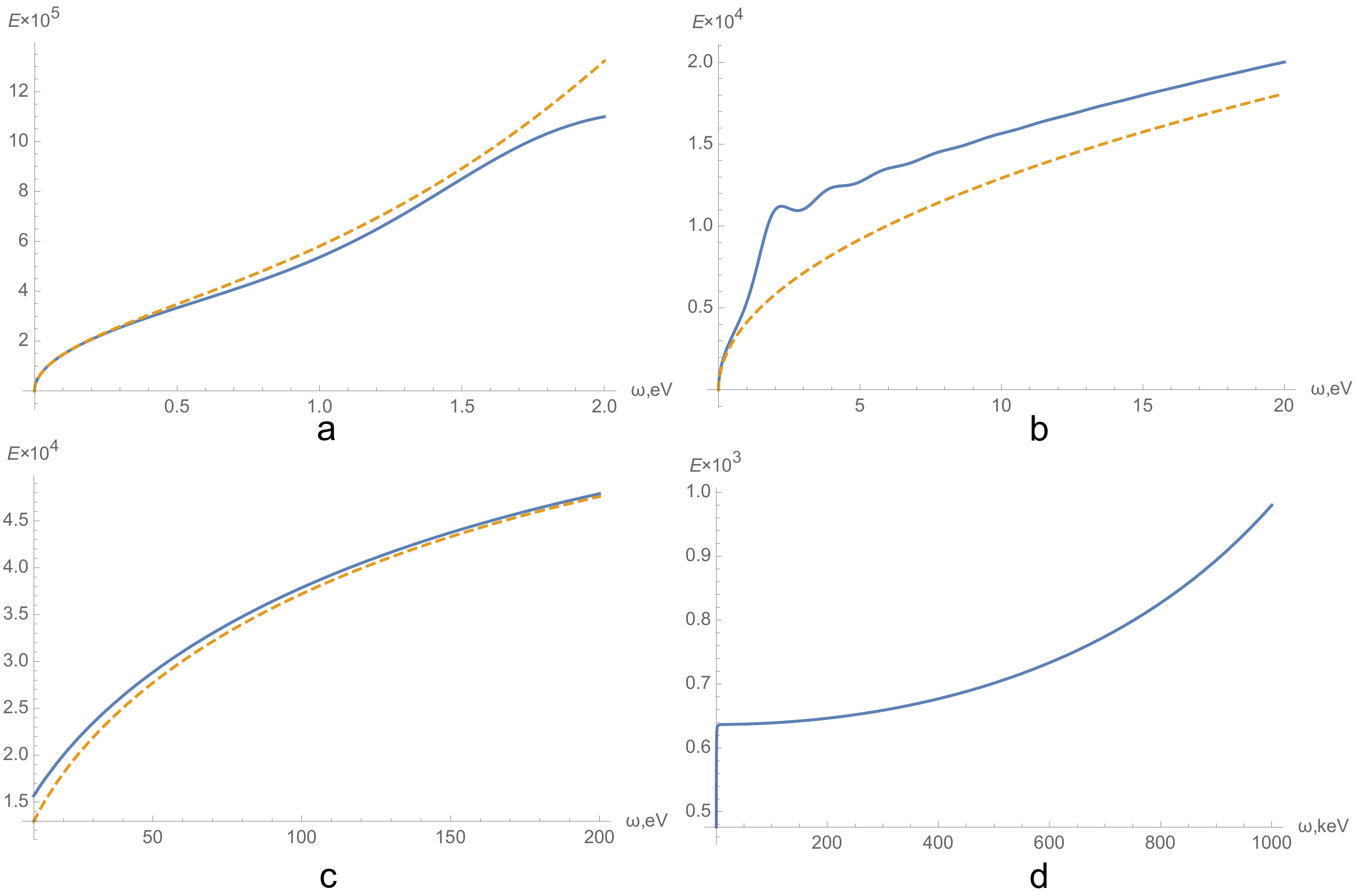}
    \caption{Quark nugget thermal emissivity in different frequency intervals. Fig.~a: solid and dashed curves represent the emissivity calculated by the  Eq.~(\ref{E-simpl}) and approximate equation (\ref{ELow}), respectively. Fig.~b: the same, but the dashed curve corresponds to the asymptotic formula (\ref{emissivity-asympt}). The oscillations of the solid curve correspond to the eigenfrequencies of the sphere with maxima at $\omega R/c =1,2,3\ldots$ Fig.~c: The same, but in larger frequency interval, where the asymptotic formula (\ref{emissivity-asympt}) provides a good accuracy. Fig.~d: emissivity for very high frequencies.}
    \label{fig-emissivity}
\end{figure*}

\begin{figure*}[htb]
    \centering
    \includegraphics[height=3cm]{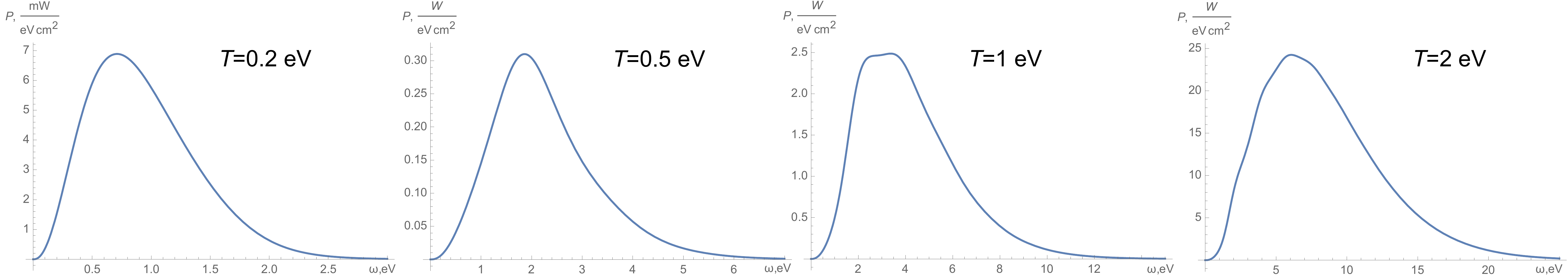}
    \caption{Radiation power from unit surface area of QN at temperatures $T=0.2$, 0.5, 1 and 2 eV.}
    \label{RadiationPower}
    \includegraphics[height=4cm]{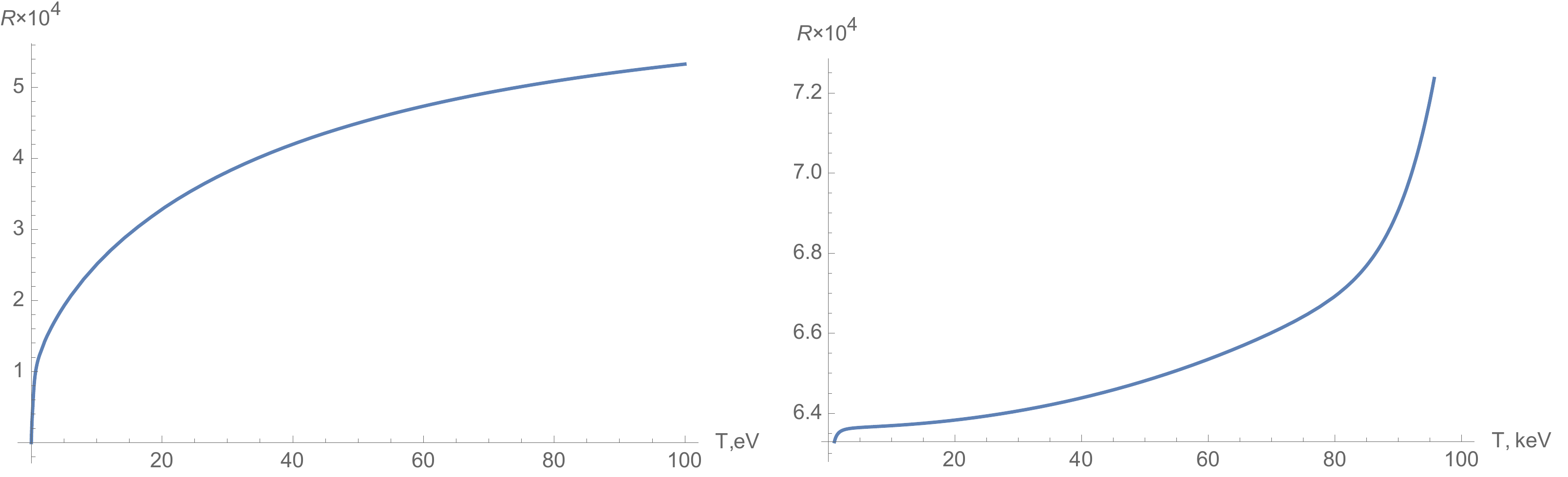}
    \caption{Ratio of the QN radiation power to the black body radiation as described by the function (\ref{R}).}
    \label{R-plots}
\end{figure*}

Making use of Eqs.~(\ref{E})--(\ref{RiccatiBessel}) and (\ref{coefficients-approx})--(\ref{emissivity-asympt}) in the regions of their applicability we calculate numerically and make plots for a range of frequencies $\omega<2$ MeV, see Fig.~\ref{fig-emissivity}. 

In Fig.~\ref{fig-emissivity}a, we plot the emissivity in the long wavelength limit, $\omega<2$ eV. The solid and dashed lines in this plot correspond to the accurate and approximate emissivites, as calculated with Eqs.~(\ref{E-simpl}) and (\ref{ELow}), respectively. 

In the intermediate frequencies, $2\mbox{ eV}\lesssim \omega \lesssim 20\mbox{ eV}$, the emissivity has oscillations combined with the rising behaviour, see
Fig.~\ref{fig-emissivity}b. The peaks of these oscillations are correlated with eigenfrequencies of the sphere, $\omega R/c=1,2,\ldots$ These oscillations have unique pattern which distinguishes the QN model from other models of dark matter. The dashed curve in Fig.~\ref{fig-emissivity}b corresponds to the asymptotic function (\ref{emissivity-asympt}) which fails to describe these oscillations and may be applied for longer wavelengths only with $\omega > 20$ eV. 

As shown in Fig.~\ref{fig-emissivity}c, for short wavelengths with $\omega> 20$ eV the oscillations corresponding to the eigenfrequencies of the sphere smooth out, and the emissivity may be well described by the asymptotic function (\ref{emissivity-asympt}). This function is represented by the dashed curve in Fig.~\ref{fig-emissivity}c, which is compared with exact expression for the emissivity (\ref{E-simpl}) given by the solid line in this plot. 

In Fig.~\ref{fig-emissivity}d we plot the emissivity of quark nugget for ultrahigh frequencies, $\omega>200$ eV. To draw this plot, the asymptotic expression for the emissivity (\ref{emissivity-asympt}) was used which is fully specified by the impedance $\zeta(\omega)$. It may be shown that the emissivity of quark nuggets as a function of frequency continues to raise up to the values $\omega=\omega_p=2$ MeV. 

\subsection{Comparison of the thermal radiation from QNs with the black body radiation}
\label{sec-comparison}

It is instructive to compare the thermal radiation from quark nuggets with the black body radiation spectrum. First, we consider the QN radiation power per unit surface area (\ref{P}). Making use the emissivity function $E(\omega)$ calculated in the previous section, we plot the QN radiation power at different temperatures in Fig.~\ref{RadiationPower}. When the temperature does not exceed a few eV, the QN radiation power has very specific pattern with visible oscillations near the eigenfrequencies of the spherical particle with $\omega R/c=1,2,3,\ldots$ This pattern, if observed, would allow to distinguish the QN model of dark matter from other ones. These oscillations smooth out and become hardly visible if the internal QN temperature exceeds 10 eV.

It is also useful to compare the total QN radiation power per unit surface area (\ref{W}) with the one of the black body (\ref{W0}). The ratio between these quantities is described by the function (\ref{R}). Making use of the QN emissivity found in Sec.~\ref{sec-emissivity}, we calculate the function (\ref{R}) numerically and plot it in Fig.~\ref{R-plots}. In Fig.~\ref{R-plots}a, we present the detailed plot for temperatures $T<100$ eV, while Fig.~\ref{R-plots}b shows the behaviour of this function at higher temperatures. These plots show that the QN radiation power is approximately by the factor $10^{-4}$--$10^{-3}$ off the one of the black body.

To further facilitate the comparison of QN radiation with the black body one, we present some values of the function (\ref{R}) in Table~\ref{TabR}.
\begin{table}[htb]
    \centering
    \begin{tabular}{c|c|c|c|c|c|c}
       $T$  & 0.1 eV & 1 eV & 10 eV & 100 eV & 1 keV & 100 keV \\\hline
     $R(T)\times10^{4}$ & 0.28 & 1.1 & 2.5 & 5.3 & 6.3 & 7.7
    \end{tabular}
    \caption{Numerical values of the function (\ref{R}) representing the ratio between the QN and black body thermal radiation power.}
    \label{TabR}
\end{table}

For $T< \gamma/10 \simeq 50$ eV the function $R(T)$ may be  described by the low-temperature analytical formula Eq.~(\ref{RT}), which for $\gamma=475$ eV and $\omega_p= 2$ MeV gives 
\begin{equation}
    R(T) =  0.87 \times 10^{-4} \sqrt{T/\mbox{eV}}\,.
    \label{RTeV}
\end{equation}

Finally, we make comparison of our results with that obtained in Ref.~\cite{WMAPhaze}. We see that our calculations give the thermal radiation power of quark nuggets by approximately 100 times bigger at the  temperature 1 eV and show different frequency and temperature dependencies. We believe that our results are more accurate for the following reasons. The authors of
Ref.~\cite{WMAPhaze} calculated the bremsstrahlung radiation of individual positrons in the positron atmosphere outside the QN core. However, a sum of radiations of individual electrons or positrons, calculated  as a two-particle problem in vacuum,  does not give correct result in a dense electron or  positron gas. When the wavelength is greater than the distance between particles,  the continuum thermal radiation is a collective effect (specific example of an important collective effect is plasma oscillations). One may say that the continuum thermal radiation is due to fluctuations of charge density. This is well known for electron gas in metals and metal nanoparticles, see, e.g., book  \cite{BHbook}. 

Ref.~\cite{WMAPhaze}  claims that radiation from  inside of QN core is absent for frequencies smaller than the plasma frequency since the radiation cannot propagate for such frequencies. Basing on this conclusion, they only calculate the bremsstrahlung radiation from a  thin positron atmosphere which contains a very small fraction of all positrons. As a result, the calculated radiation is very small. However, we know that heated metals radiate and frequency of their thermal radiation is typically much smaller than the plasma frequency. Fluctuating electromagnetic field, induced by the  fluctuations of the charge density, does not  vanish inside. Actually, it may be much bigger than the field  in the radiated waves, see, e.g., book  \cite{BHbook}. However, only a distant tail of this field may be identified  with the radiation field which is well defined  on the distance from the source bigger than the photon wavelength. 
 
For the radiation  with photon energy $\sim 10^{-4}$ eV, studied in Ref. \cite{WMAPhaze} to explain the ``WMAP haze'', the vacuum wavelength  $\lambda\sim 10^{8}\, a_B$, where $a_B$ is the Bohr radius. It is 10 orders of magnitude bigger than the width of the positron atmosphere. To avoid misunderstanding, inside QN the wavelength is not defined since the photon mean free path is very small. One may talk about external photon penetration length which is still much bigger than the positron atmosphere width.

Note  that a metal nanoparticle has electron ``atmosphere'' of a bigger  size than that  of QN, about few  $a_B$. For QN the atmosphere width (where the positron density decreases 8 times)  is $z_0\sim 0.003 a_B$ \cite{WMAPhaze}. Therefore, one  should consider models accounting for the total radiation of QN, similar to the radiation of metal nanoparticles.


\section{Thermal equilibrium in different media}
\label{Equilibrium}

The quark nugget model of dark matter suggests that the dark matter is composed of  quark and anti-quark nuggets, with the latter approximately 1.5 times more abundant in the universe than the former. The anti-quark nuggets are of special importance in this model because they are built of antimatter: they have antiquark core surrounded by the positron cloud. In this section, we focus on the properties of antiquark nuggets because they effectively interact with visible matter. As is demonstrated in Refs.~\cite{Electrosphere2008,FS21}, their annihilation cross section in collisions with atoms and molecules of visible matter is close to its geometric cross section. Protons and atomic nuclei annihilate with antiquarks in the quark core with the emission of energetic $\pi$ mesons. About 50\% of the annihilation products penetrate deeper into the quark nugget and their energy is thermalized. Therefore, we can roughly estimate that each hydrogen atom colliding with the quark nugget contributes 1 GeV of thermal energy. In the following subsections we will calculate the internal temperature of antiquark nuggets in the thermal equilibrium in different media.

\subsection{Radiation from QNs in our galaxy}
\label{sec-equilib-GC}

In Ref.~\cite{WMAPhaze} it was conjectured that thermal radiation from quark nuggets may be responsible for the diffuse radiation in the center of our galaxy registered by the WMAP observatory and referred to as the ``WMAP hase'' \cite{WMAP1,WMAP2,WMAP3,WMAP4}. Therefore, it is important to consider the condition of thermal equilibrium for (anti)quark nuggets in the interstellar medium in the center of our galaxy.

\subsubsection{QN effective temperature in galactic bulge}

The power per unit surface area of thermal radiation of a quark nugget is given by Eq.~(\ref{WR}). Thus, the total radiation power is $W_{\rm rad}=4\pi R^2 W(T)$. This emission should be balanced by the flux of incoming energy due to collisions of antiquark nuggets with the visible matter. The average density of gas in the interstellar medium near the galaxy center is $n_{\rm vm} = 8.2\mbox{ cm}^{-3}$ \cite{gas}. Assuming that the average velocity of dark matter particles through the visible matter is of order $v=10^{-3}c$, and that each nucleon colliding with QN brings up about 1 GeV of energy, we estimate the incoming energy flux per unit area as
\begin{equation}
    W_{\rm in} = 1\mbox{ GeV} \sigma_{\rm ann} v n_{\rm vm}\,,
    \label{Win}
\end{equation}
where $\sigma_{\rm ann}$ is the annihilation cross section for a nucleon colliding with QN. In Ref.~\cite{FS21}, it was argued that $\sigma_{\rm ann}\approx \pi R^2$, so the equilibrium temperature is found from the equation $W_{\rm in}=W_{\rm rad}$, which implies
\begin{equation}
   4 R(T)\sigma T^4 = 1\mbox{ GeV} v n_{\rm vm}\,.
   \label{equilib}
\end{equation}
Given the function $R(T)$ from Fig.~\ref{R-plots}a, or its interpolation from values in Table~\ref{TabR}, we find
\begin{equation}
    T = 0.2\mbox{ eV}.
    \label{temp}
\end{equation}

We point out that the temperature estimate (\ref{temp}) is independent of QN baryon charge number $B$ (and the QN radius), because the radius $R$ cancelled out in Eq.~(\ref{equilib}).

We should note a very weak dependence of the QN temperature on the parameters of the problem. Using the low temperature formula Eq. (\ref{RTeV}) for $R(T)$ we can cast  Eq.~(\ref{equilib}) in the following form:
\begin{equation}
   32\frac{\sqrt{T \gamma}}{\omega_p} \sigma T^4 =  W_{\rm in}=1\mbox{ GeV} v n_{\rm vm}\,.
   \label{equilib1}
\end{equation}
This gives 
\begin{equation}
 T=  ( W_{\rm in}/32\sigma)^{2/9} (\omega_p)^{2/9} (\gamma)^{-1/9}.
   \label{Tequilib}
\end{equation}
This formula is valid for $T<\gamma/10\simeq50$ eV. However, for higher temperature $R(T)$ varies very  slowly (see Table \ref{TabR}) and  the dependence on the  parameters continues to be weak, $T = ( W_{\rm in}/4 \sigma R(T))^{1/4}$. 

Note that here we have taken into account only interaction of quark nuggets with the gas in the interstellar medium and neglected interaction with stars because we are interested only in the diffuse radiation in our galaxy. The diffuse radiation appears mainly in the infrared spectrum, with maximum of intensity at $\omega\approx 0.7$ eV (see Fig.~\ref{RadiationPower}). 

\subsubsection{QN effective temperature in a dense molecular cloud}

A typical giant molecular cloud has a diameter of order 400 pc and hydrogen molecule number density 300 cm$^{-3}$ \cite{AstrophysicsBook2}. Thus, the nucleon number density in such cloud is $n_{\rm cloud} = 600\mbox{ cm}^{-3}$. Substituting this density into Eq.~(\ref{equilib1}) instead of $n_{\rm vm}$, we solve for the QN effective temperature in the molecular cloud,
\begin{equation}
    T = 0.5\mbox{ eV}.
    \label{Tcloud}
\end{equation}
This temperature is significantly higher than the QN effective temperature (\ref{temp}) estimated with average gas density in the galactic bulge. The maximum of the radiation intensity is in visible light spectrum and QN radiation may be observed by optical telescopes, see Fig.~\ref{RadiationPower}. It would be interesting to study the radiation in the visible spectrum from dense molecular clouds in different areas of our galaxy. The intensity of this radiation should correlate with dark matter density distribution.

\subsubsection{QN effective temperature in the Sun neighborhood}

It is useful to estimate the effective QN temperature in the interstellar medium in the solar neighbourhood, where the gas number density is $n_{\rm vm} \approx 1.6 \mbox{ cm}^{-3}$ \cite{AstrophysicsBook}. Solving Eq.~(\ref{equilib}) with this density would yield
\begin{equation}
    T_{\rm sun} =0.15\mbox{ eV}.
\end{equation}
Thus, the QN effective temperature varies slowly with the distance from the galactic center.

\subsection{Radiation from QNs in the air}

In Ref.~\cite{BFZ} it was argued that quark nuggets crossing the Earth's atmosphere can create specific acoustic waves which may be discriminated from background and detected.
QN radiation may also be detected by radars Ref.~\cite{BFZ}.
Therefore, it is important to determine the QN temperature in the air. 

To get an upper estimate of the temperature, we consider the air density at the ground level $n_{\rm air} = 7.3\times 10^{20}$ nucleons/cm$^3$, and substitute $n_{\rm vm }\to n_{\rm air}$ in Eq.~(\ref{Win}). Then, the thermal equilibrium condition
$W_{\rm in} = W_{\rm rad}$ implies
\begin{equation}
    T = 11\mbox{ keV}.
\end{equation}
At higher altitudes with thinner air this temperature would be lower.
Note that this result is close to  the QN temperature estimate in Ref.~\cite{BFZ}.

\subsection{Radiation from QNs crossing the Earth}

In Ref.~\cite{Ge2020} it was argued that the quark nuggets crossing the Earth produce a specific radiation which may be observed by the  XMM-Newton satellite \cite{XMMN}. 
High intensity QN radiation  also rapidly heats environment and produces shock/specific acoustic waves which may be discriminated from seismic background and detected Ref.~\cite{BFZ}.
Therefore it is interesting to estimate the QN equilibrium temperature when it crosses dense celestial bodies like the Earth.

With the average density of the Earth $n_{\rm earth} = 3.3\times 10^{24}$ nucleons/cm$^3$, we solve the equilibrium equation $W_{\rm in} = W_{\rm rad}$. Given the function $R(T)$ in Fig.~\ref{R-plots}b, or its extrapolation from values in the Table~\ref{TabR}, we find
\begin{equation}
    T = 87\mbox{ keV}.
\end{equation}
This temperature is a few times lower than the corresponding estimate in Ref.~\cite{XMMN}. This difference may be explained by the underestimation of the QN cooling rate in the model considered in Ref.~\cite{XMMN}.

\section{Manifestation of QN radiation in galaxy}
\label{Galaxy}

In this section, we revisit some of the QN detection proposals based on the thermal radiation pattern of quark nuggets determined in this paper. 

\subsection{Can QNs explain WMAP haze?}

WMAP detected an excess of microwave radiation from the center of our galaxy \cite{WMAP1,WMAP2,WMAP3,WMAP4}. In Ref~\cite{WMAPhaze} it was argued that the axion-quark nuggets can emit the radiation in our galaxy, which was detected by WMAP. In this section, we verify this hypothesis in the light of QN thermal radiation pattern found in Sect.~\ref{sec-radiation}.

According to Ref.~\cite{WMAP5}, the WMAP registered an excess of radiation at frequencies 22, 33, 41, 61 and 93 GHz around the galactic center in the angular apertures of 20\textdegree. The measured energy flux is 
\begin{equation}
    \Phi_{\rm haze} = \mbox{1-5 kJy sr}^{-1}=
     \frac{\mbox{(1-5)}\times 10^{-20} \rm erg}{\mbox{sec cm$^2$ Hz sr}}
    \,.
    \label{observed-haze}
\end{equation}
It was conjectured in Ref.~\cite{WMAP5} that this diffuse radiation around the center of our galaxy may be produced by annihilation of WIMPs. Here we will check if this radiation may be attributed to QNs.

\subsubsection{Thermal emission at GHz frequencies}

Recall that Eq.~(\ref{P}) gives the radiation power from a unit surface element $ds$ of a spherical particle with surface emissivity $E(\omega)$ in the frequency interval $(\omega,\omega+d\omega)$. This radiation power is obtained by integrating over all possible directions as in Eq.~(\ref{P0}). In a similar way one finds the energy flux from this surface element only in the direction normal to $ds$,
\begin{equation}
I(\omega,T) = E(\omega) I_0(\omega,T)\,,
\label{I}
\end{equation}
where $I_0(\omega,T)$ is the Planck function (\ref{I0}). The function (\ref{I}) quantifies the energy flux of thermal radiation per unit time from a surface element $ds$ of a spherical particle in the direction normal to $ds$ in a unit solid angle in a frequency interval $(\omega,\omega+d\omega)$. The total energy flux per unit time from the spherical particle in any given direction may be found by integrating the function (\ref{I}) over a half of the sphere $S/2$ with radius $R$,
\begin{equation}
    F (\omega,T) = \int_{S/2} \vec I(\omega,T)\cdot d\vec s = \frac12 \pi^2 R^2 I(\omega,T)\,.
    \label{F}
\end{equation}

The mass of each QN may be estimated as $B$\,GeV, with $B$ being the baryon charge number. Therefore, in the QN model of dark matter, the dark matter particle number density may be written as
\begin{equation}
\label{nDM}
    n_{\rm DM} = \frac{  \rho_{\rm DM} }{B\mbox{ GeV}}\,,
\end{equation}
where $\rho_{\rm DM}$ is the dark matter density.

There is no unique choice of the dark matter density profile in our galaxy. Following Ref.~\cite{WMAP5}, we will use the density profile which scales as $\rho_{\rm DM}(r)\propto r^{-1.2}$ near the galactic center. More specifically, the dark matter density may be described by
\begin{equation}
    \label{rhoDM}
    \rho_{\rm DM}(r) = \frac{2.4\times 10^8 M_\odot }{r^{1.2}(1+r/R_s)^{1.8} \mbox{kpc}^{1.8}}\,,
\end{equation}
with $R_s = 20$~kpc. This dark matter density is correctly normalized to reproduce the local dark matter density $\rho= 0.4$ GeV/cm$^3$. A similar profile was used in the recent paper \cite{FS21} for an estimate of a flux of 511 keV photons produced in the process of electron-positron annihilations in the QNs.

Given the energy flux of one QN particle (\ref{F}) we find the corresponding energy flux per unit surface area per unit solid angle at the observation point on Earth,
\begin{equation}
    \Phi_{\rm QN} = \int_\ell F(\omega,T) n_{\rm DM} d\ell
    =\frac{\pi^2 R^2}{2B \mbox{ GeV}}
    \int_\ell I(\omega,T) \rho_{\rm DM} d\ell
    \,,
    \label{Phi}
\end{equation}
where the integration is performed along the line of sight $\ell$. Here we have made use of Eqs.~(\ref{F}) and (\ref{nDM}).

In Sect.~\ref{sec-equilib-GC} it was shown that the QN effective temperature varies slowly with the distance from the galactic center. Therefore, in the leading-order approximation the function $I(\omega,T)$ in Eq.~(\ref{Phi}) may be considered as a constant,
\begin{equation}
    \Phi_{\rm QN} 
    \approx\frac{\pi^2 {\rm fm}^2I(\omega,T)}{2B^{1/3} \mbox{ GeV}}
    \int_\ell \rho_{\rm DM} d\ell
    \,,
    \label{Phi1}
\end{equation}
where we made use of the expression for the QN radius in terms of the baryon number (\ref{radius-relation}).

The value of the integral $\int_\ell \rho_{\rm DM} d\ell$ depends on the choice of the line of sight. We consider the lines of sight at the galactic longitudes from 6\textdegree\ to 14\textdegree\ which correspond to the observation directions of WMAP \cite{WMAP5}. For these lines of sight, using the dark matter density profile (\ref{rhoDM}), we find
\begin{equation}
    \int_\ell \rho_{\rm DM} d\ell =(\mbox{0.7-1.2})\times 10^{-3} \mbox{GeV fm}^{-2}\,.
    \label{LOSint}
\end{equation}

The energy flux through the unit surface area of a quark nugget (\ref{I}) can be calculated with the help of Eq.~(\ref{ELow}) since the WMAP haze frequency 33 GHz corresponds to the long-wavelength regime with $\omega = 10^{-4}$ eV. For the equilibrium temperature $T=0.2$ eV we find
\begin{equation}
    I= 3\times 10^{-20} \frac{\rm erg}{\mbox{s cm}^2\mbox{ Hz sr}}\,.
    \label{I-res}
\end{equation}
Substituting Eqs.~(\ref{LOSint}) and (\ref{I-res}) into (\ref{Phi}) we find
\begin{equation}
    \Phi_{\rm QN} \approx \frac{(\mbox{1-1.8})\times 10^{-22} }{B^{1/3}}
    \frac{\rm erg}{\mbox{s cm}^2\mbox{ Hz sr}}\,.
\end{equation} 
This flux is significantly lower than the observed value (\ref{observed-haze}) for any value of the baryon charge number. Also, the spectrum of QN thermal radiation looks different to that observed.  Thus we conclude that the observed excess of the diffuse microwave radiation in our galaxy termed as the {\it WMAP haze cannot be explained by the thermal radiation of quark nuggets}. 

The quark nuggets, however, may produce other types of radiation, which are not attributed to the thermal emission. In particular, they can emit $\pi$ mesons originating from the annihilation events in collisions with the gas particles in the interstellar medium.

\subsubsection{Synchrotron radiation at GHz frequencies}
\label{SecSynch}

In Ref.~\cite{WMAP5} it was argued that WIMPs may produce the radiation similar to those identified as the WMAP haze in Refs.~\cite{WMAP1,WMAP2,WMAP3,WMAP4} if the WIMP mass is of order 100 GeV. Indeed, WIMPs may decay into energetic electron-positron pairs, which would emit synchrotron radiation upon interaction with magnetic fields in the galaxy. For magnetic fields of order $H=10\, \mu$G this synchrotron radiation appears comparable with the one detected by WMAP \cite{WMAP5}. In this section, we check if this radiation may originate from quark nuggets.

As is shown in Ref.~\cite{FS21}, the annihilation cross section of anti-QNs with hydrogen and helium gases is close to the geometric cross section $\pi R^2$. The annihilation of protons in the anti-quark core leads to emission of $\pi$ mesons, similar to the nucleon-antinucleon annihilation process. On average, about five $\pi$ mesons are produced in such annihilation, of which about one pair is given by the charged ones. These $\pi^\pm$ mesons undergo a chain of decay reactions with $e^+$ and $e^-$ (as well as neutrinos and photons) arising in the final products. Roughly one half of these particles are emitted outside the QNs while the others thermalize deep inside the quark core and further decay. Thus, each collision of a hydrogen atom with the quark nugget produces at least one electron or positron with energy of order 100-400 MeV. Can these electrons produce a significant amount of synchrotron radiation upon interaction with the weak magnetic field in the galaxy?

The spectral density of synchrotron radiation is given by \cite{LL2}
\begin{equation}
    I(\omega) = \frac{\sqrt3 }{2\pi}
    \frac{e^3 H}{mc^2} F(\omega/\omega_c)\,,
    \label{synch-radiation}
\end{equation}
where
\begin{equation}
    F(x) = x\int_x^\infty K_{5/3}(z)dz\,,
    \quad
    \omega_c = \frac{3eH}{2mc}\left(
    \frac{{\cal E}}{mc^2}
    \right)^2\,.
    \label{FK}
\end{equation}
Here $K_{5/3}(z)$ is the modified Bessel function of the second kind and $\cal E$ is the particle's energy. An electron with kinetic energy ${\cal E}=400$ MeV in the magnetic field $H=10\,\mu$G has the frequency $\omega_c\approx 10^{-7}$ eV, and for the observed frequencies $\omega=10^{-4}$ eV the argument of the function $F(x)$ in Eq.~(\ref{synch-radiation}) is large, $x=\omega/\omega_c \approx 10^3$. For large arguments this function has the following asymptotic behaviour $F(x)\simeq \sqrt{\pi x/2}e^{-x}$ \cite{AstrophysicsBook1}. Thus, the frequencies $\omega=10^{-4}$ eV in the synchrotron radiation from emitted electrons or positrons is strongly suppressed by the factor $\sim e^{-1000}$, which makes it unobservable. We conclude that unlike the WIMP decay scenario suggested in Ref.~\cite{WMAP5}, synchrotron radiation of electrons and positrons emitted from QNs {\it cannot explain the WMAP haze}.

Note that the above suppression of radiation persists at the frequencies well above $\omega_c$, $\omega\gg \omega_c$.
In Sect.~\ref{C1} we will estimate the maximum of this synchrotron radiation at $\omega\sim\omega_c$ and study its contribution to the galaxy RF backround radiation.

\subsubsection{Bremsstrahlung radiation}

Annihilation of gases in the interstellar medium with antiquark nuggets yields the emission of $\pi^\pm$ mesons, similar to the nucleon-antinucleon annihilation \cite{FS21}. These $\pi^\pm$ mesons have typical energies 100-400 MeV. Therefore, they may produce bremsstrahlung radiation when they move through the positron cloud of antiquark nugget. 

Consider a positron with energy $\varepsilon_1$ scattering off a Coulomb potential of a charged $\pi$ meson with emission of a proton with energy $\hbar\omega'$. In the $\pi$ meson rest frame, the differential cross section of this process reads \cite{AB}
\begin{equation}
\label{brems-sross-section}
\begin{split}
    \frac{d\sigma}{d\omega'} &= \frac{\alpha r_0^2}{\omega'}\frac{p_2}{p_1} \bigg\{
    \frac43 - 2\varepsilon_1\varepsilon_2 \frac{p_1^2 + p_2^2}{p_1^2 p_2^2} + \\
    &+m^2 \left(
    \frac{\eta_1\varepsilon_2}{p_1^3}
    +\frac{\eta_2\varepsilon_1}{p_2^3}
    -\frac{\eta_1 \eta_2}{p_1 p_2}
    \right) \\
    &+L\bigg[
    \frac83 \frac{\varepsilon_1\varepsilon_2}{p_1p_2}
    +\frac{(\omega')^2}{p_1^3 p_2^3}(\varepsilon_1^2 \varepsilon_2^2 +p_1^2 p_2^2)
    \\
    &+\frac{m^2 \omega'}{2p_1 p_2}
    \left( \eta_1 \frac{\varepsilon_1 \varepsilon_2+p_1^2}{p_1^3}
     -\eta_2\frac{\varepsilon_1\varepsilon_2 + p_2^2}{p_2^3}
     +\frac{2\omega'\varepsilon_1\varepsilon_2}{p_1^2 p_2^2}\right)
     \bigg]\bigg\},
\end{split}
\end{equation}
where
\begin{eqnarray}
L&=& 2\ln\frac{\varepsilon_1 \varepsilon_2 + p_1 p_2 - m^2}{m\omega'}\,,\\
\eta_1&=& 2\ln\frac{\varepsilon_1 + p_1}{m}\,,\quad
\eta_2= 2\ln\frac{\varepsilon_2 + p_2}{m}\,.
\end{eqnarray}
Here $\alpha\approx \frac1{137}$ is the fine structure constant, $r_0 = \frac{e^2}{mc^2}$ is the classical radius of electron. The energy $\varepsilon_2$ and momentum $p_2$ of the outgoing positron may be expressed via the corresponding quantities of the incident positron, $\varepsilon_2 = \varepsilon_1 - \hbar\omega$, $p_2^2 = (\varepsilon_1 - \hbar\omega)^2 - m^2$.

Assume, for simplicity, that in the QN rest frame the $\pi$ meson moves along the $z$ axis with velocity ${\bf v}_\pi = (0,0,v_\pi)$, $v_\pi\approx 0.94 c$. Positron and photon energies and momenta in Eq.~(\ref{brems-sross-section}) should be Lorentz transformed to the QN rest frame, $(\varepsilon_1,{\bf p}_1)\to (\varepsilon,{\bf p})$, $\omega'\to\omega$. This will result in quite complicated but explicit expression for the cross section in the QN rest frame, $d\sigma(p)/d\omega$. Effectively, this cross section is a function of the incoming positron momentum $p$ only.

Radiation power per unit frequency of the bremsstrahlung radiation is given by
\begin{equation}
\label{PBS}
    P = \left\langle n_{e^+} v \omega \frac{d\sigma}{d\omega}\right\rangle \,,
\end{equation}
where $n_{e^+}v =n_{e^+} p/\varepsilon$ is the positron current density at the position of the $\pi$ meson. In Eq.~(\ref{PBS}), we have to average over the positron momenta $\bf p$ inside the Fermi sphere, $|{\bf p}|<p_F$,
\begin{equation}
\label{P-averaged}
    P = \frac{2}{(2\pi)^3} \int_{|{\bf p}|<p_F}\omega
    \frac{d\sigma(p)}{d\omega} \frac{pd^3p}{\sqrt{p^2 + m^2}}\,.
\end{equation}
Thus, the radiation power is a function of the Fermi momentum of positrons in the positron cloud, $P=P(p_F)$, $p_F^2 = ({\cal E}_F + m)^2 - m^2$. Here, the Fermi energy ${\cal E}_F$ may be identified with the positron chemical potential within the Thomas-Fermi model, ${\cal E}_F =\mu$, see \cite{FS21}. 

Near the boundary of the quark core the positron chemical potential $\mu$ varies rapidly \cite{FS21}. The total energy per unit frequency produced by the bremsstrahlung radiation from one antiquark nugget is given by the integral of the radiation power (\ref{P-averaged}) over the path $\ell$ of the $\pi$ meson in the positron cloud,
\begin{equation}
\label{Eb}
    {\cal E}_{\rm brems}= \frac1{v_{\pi}} \int_\ell d\ell\, P\,.
\end{equation}
To get the radiation power from all events of hydrogen annihilation on quark nuggets in the bulge of our galaxy, we multiply the radiation power (\ref{Eb}) by the estimated below hydrogen annihilation rate (\ref{F-rate}),
$F{\cal E}_{\rm brems}$. At the observation point on Earth, the corresponding photon flux would be
\begin{equation}
    P_{\rm brems} = \frac{F{\cal E}_{\rm brems}}{\Theta S}\,,
\end{equation}
where $\Theta=0.16$ sr is a visible solid angle of bulge of our galaxy and $S$ is the area of the sphere with radius $8.5$ kpc around the galaxy center.

The integration in Eqs.~(\ref{P-averaged}) and (\ref{Eb}) may be performed numerically for each particular photon frequency. For $\omega=10^{-4}$ eV, we find
\begin{equation}
    P_{\rm brems} = 1.4\times 10^{-24} B^{-1/3}\frac{\rm erg}{\mbox{s cm}^2\,\rm Hz\ sr}\,.
\end{equation}
This radiation is significantly lower than the observed WMAP haze flux (\ref{observed-haze}). Thus, the {\it bremsstrahlung radiation from quark nuggets in the center of our galaxy cannot explain the WMAP haze} for any value of the baryon charge $B$. 


\subsection{Can QNs be responsible for IR diffuse radiation?}

As is shown in Sect.~\ref{sec-equilib-GC}, the effective temperature of QNs in the galaxy is about $T=0.2$ eV. According to Fig.~\ref{RadiationPower}, the maximum of radiation from QNs at this temperature corresponds to the near-IR frequencies $\omega = 0.5-1$ eV. Can quark nuggets produce a significant portion of IR radiation in our galaxy, which might be detected?

To answer this question, we should compare the QN radiation with the infrared background radiation in our galaxy observed by Cosmic Background Explorer (COBE) satellite. In particular, at the frequency $\omega = 0.56$ eV ($\lambda = 2.2\ \mu$m) it registered the radiation flux \cite{COBE}
\begin{equation}
\label{PhiCOBE}
\begin{split}
    \Phi_{\rm COBE} &= 15-100 \mbox{ MJy sr}^{-1} 
    \\&=  (1.5-10)\times 10^{-16} \mbox{ erg s}^{-1} {\rm cm}^{-2}{\rm Hz}^{-1}{\rm sr}^{-1}
\end{split}
\end{equation}
at the galactic latitudes $|b|\leq3\degree$. For simplicity, we will consider further the latitude $b=3\degree$ with the IR radiation flux $\Phi_{\rm COBE}=1.5\times 10^{-16} \mbox{ erg s}^{-1} {\rm cm}^{-2}{\rm Hz}^{-1}{\rm sr}^{-1}$.

The quark nugget surface emissivity at the same frequency may be found from Eq.~(\ref{I}),
\begin{equation}
    I = 1.3\times 10^{-11} \mbox{ erg s}^{-1}{\rm cm}^{-2}{\rm Hz}^{-1}{\rm sr}^{-1}\,.
\end{equation}
This quantity should be substituted into Eq.~(\ref{Phi1}) together with the line-of-sight integral corresponding to the galactic latitude $b=3\degree$,
\begin{equation}
    \int_\ell \rho_{\rm DM} d\ell =1.7\times 10^{-3} \mbox{GeV fm}^{-2}\,.
    \label{LOSint1}
\end{equation}
As a result, we find the flux of IR radiation from QNs produced in our galaxy
\begin{equation}
    \Phi_{\rm QN} \approx \frac{ 10^{-13} }{B^{1/3}}
    \frac{\rm erg}{\mbox{s cm}^2\mbox{ Hz sr}}\,.
\end{equation} 
This flux becomes comparable with (\ref{PhiCOBE}) if
\begin{equation}
    B<3.8\times 10^{8}\,.
\end{equation}
Unfortunately, this constraint is incompatible with the earlier established bound (\ref{Bconstraint}). Thus, {\it thermal radiation from QNs gives negligible contribution to the galaxy IR background radiation.}

\subsection{RF diffuse radiation from QNs}

In Sect.~\ref{SecSynch} we considered the synchrotron radiation from ultrarelativistic electrons and positrons emitted from antiquark nuggets in the process of annihilation with hydrogen atoms and molecules in the galaxy. These electrons emit the synchrotron radiation when they move in the random magnetic fields with characteristic magnitude of order $H\sim 10\ \mu$G. As we demonstrated, this radiation is strongly suppressed at GHz frequencies, but it is maximized at RF frequencies in the MHz range. In this section, we consider the synchrotron radiation in the MHz range and compare it with the galaxy RF background radiation.

\subsubsection{Synchrotron radiation at MHz frequencies}
\label{C1}

The function (\ref{FK}) reaches its maximum $F_{\rm max}\approx 0.92$ at $x_{\rm max} = 0.29$. This maximum corresponds to radiofrequencies
\begin{equation}
\omega = 0.29 \omega_c = 0.29\times 10^{-7}\mbox{ eV} 
=44\mbox{ MHz}\,.
\label{RF}
\end{equation}
For this frequency, the spectral density (\ref{synch-radiation}) may be estimated as
\begin{equation}
    I(\omega_{\rm max}) = \frac{\sqrt3 }{2\pi}
    \frac{e^3 H}{mc^2} F_{\rm max} \approx
    3.4 \times 10^{-28}\mbox{ erg s}^{-1}{\rm Hz}^{-1}\,.
\end{equation}
This radiation should be distributed diffusely with a peak around the central bulge of our galaxy.

The annihilation rate per unit volume of the hydrogen gas in collisions with the antiquark nuggets in our galaxy is 
\begin{equation}
    \label{ann-rate}
    W = \sigma v n_{\rm DM} n_{\rm H}
    \approx \pi v B^{-1/3} \rho_{\rm DM} n_{\rm H} 
    \frac{{\rm fm}^2}{\rm GeV}
    \,,
\end{equation}
where $\sigma\approx \pi R^2=\pi B^{2/3}(1\,{\rm fm})^2$ is the annihilation cross section, $v=10^{-3}c$ is the average velocity of dark matter particles, $n_{\rm DM}$ is the dark matter particle density (\ref{nDM}) and $n_{\rm H}$ is the hydrogen gas density in the galaxy. The latter may be roughly modelled by a spherically symmetric distribution around the center of our galaxy \cite{gas}
\begin{equation}
\label{gas-density}
    n_{\rm H}(r) = n_0 \exp(-r/h)\,,
\end{equation}
where\footnote{This value for $n_0$ is twice the value for $\rho_{H_2}$ given in Table 4 in Ref.~\cite{gas}, because each $H_2$ molecule contributes two protons which annihilate with the antiquark core. Note also that we neglect the contribution of the atomic hydrogen as its density in the galactic bulge is much lower than that  of the molecular hydrogen.} 
$n_0=8.1\mbox{ cm}^{-3}$ and $h=2.57$ kpc.

Making use of the dark matter (\ref{rhoDM}) and hydrogen gas (\ref{gas-density}) densities, we integrate the annihilation rate (\ref{ann-rate}) over the bulge of our galaxy, which may be roughly considered as a spherical region with radius $R_{\rm bulge} \approx2$ kpc around the galactic center:
\begin{equation}
\label{F-rate}
    F= \int_{\rm bulge} W d^3r = 2.8\times 10^{49} B^{-1/3} {\rm s}^{-1}\,.
\end{equation}
This equation describes the production rate of ultrarelativistic electrons or positrons in the process of annihilation of hydrogen gas with anti-quark nuggets in the bulge of our galaxy. How much time do these particles need to leave the bulge? Relativistic cosmic rays in the galaxy have random walk trajectories along the lines of magnetic fields \cite{CosmicRays}. These magnetic fields have a characteristic correlation length $l=100$ pc. The diffusion time of an ultrarelativistic particle in the bulge of the galaxy may be roughly estimated as $t=R_{\rm bulge}^2/(l c)\sim 10^5$ yr. However, the electrons may be trapped in the magnetic fields for significantly longer \cite{CosmicRays}. This may increase the diffusion time of the electron in the galactic bulge up to $t=10^6$ yr. 
Thus, very approximately, at any given moment of time there are about 
\begin{equation}
 N= Ft = 10^{63} B^{-1/3} 
\end{equation}
ultrarelativistic electrons and positrons in the bulge produced by anti-quark nuggets. The spectral density from all these electrons at the frequency (\ref{RF}) is
\begin{equation}
\label{61}
    N I(\omega_{\rm max}) = 4\times 10^{35} B^{-1/3}
    \mbox{ erg s}^{-1}{\rm Hz}^{-1}\,.
\end{equation}
To get the corresponding radiation spectral density at the observation point on Earth we have to divide the result (\ref{61}) by the area of the sphere $S$ with the radius $8.5$ kpc and the value of the solid angle at which the galaxy bulge is visible from Earth $\Theta$,
\begin{equation}
   P_{\rm QN}= \frac{N I(\omega_{\rm max})}{\Theta S}
    =\frac{2.7\times 10^{-10}}{ B^{1/3}}
    \frac{\rm erg}{\mbox{s cm}^{2}\,\rm Hz \ sr}\,,
    \label{flux-calculated}
\end{equation}
where
\begin{equation}
\label{ThetaS}
    S = \pi (8.5\mbox{ kpc})^2\,,\qquad
    \Theta = 0.16\mbox{ sr}\,.
\end{equation}

The flux (\ref{flux-calculated}) may be compared with the radiofrequency diffuse galactic radiation which has a synchrotron background radiation origin as well \cite{RF1,RF}. At the 40 MHz frequency, this flux is of order
\begin{equation}
P_{\rm backg}\approx 3 \times 10^{-21}\frac{\rm W}{{\rm m}^2\, \rm Hz\ sr} = 3\times 10^{-18}\frac{\rm erg}{\mbox{s cm}^2\,\rm Hz\ sr}\,.
\label{flux-observed}
\end{equation}
Comparing Eq.~(\ref{flux-calculated}) with (\ref{flux-observed}), we conclude that the radiation from quark nuggets may be resolved from the background if
\begin{equation}
    B< 7.6\times 10^{23}\,.
\end{equation}
This result is very close to the bound (\ref{Bconstraint}). Thus, {\it it is plausible that quark nuggets may contribute significantly} to the background radiation from the galaxy bulge at MHz frequencies.


\subsection{Can QNs be responsible for diffuse $\gamma$-ray radiation?}

The annihilation of hydrogen atoms in the antiquark core yield the emission of $\pi$ mesons. This annihilation happens near the surface of the QN core, at the depth of about a one fm \cite{FS21}. Neutral $\pi^0$ mesons decay into $\gamma$ photons with energies of order 200-300 MeV, which contribute to the galaxy $\gamma$ ray background radiation.
Moreover, when charged $\pi^\pm$ mesons move through the positron cloud and then leave the QN through the surface, they may produce the transition radiation.

\subsubsection{$\pi^0$ meson decay}
On average, each hydrogen annihilation event on the anti-QN yields two $\pi^0$ mesons, each has the main decay channel into two $\gamma$ photons with an energy of order $E_\gamma\sim 200$ MeV. The hydrogen annihilation rate in the galaxy is given by Eq.~(\ref{ann-rate}). By integrating this annihilation rate over the line of sight $\ell$ and dividing by the full solid angle $4\pi$ we determine the flux of $\gamma$ photons from decaying $\pi^0$ mesons at the observation point on Earth (number of photons per second per unit surface area per steradian),
\begin{equation}
    F = \frac{4}{4\pi} \int_\ell W d\ell\,.
    \label{gamma-flux}
\end{equation}
The factor 4 in the numerator takes into account four emitted photons from each annihilation event. 

Upon averaging Eq.~(\ref{gamma-flux}) over the lines of sight in the sector with the galactic coordinates $|l| < 80^\circ$, $|b| < 8^\circ$, we find the mean photon flux with typical energy $200$ MeV,
\begin{equation}
    \langle F\rangle = 1.4 \times 10^{5} B^{-1/3}\mbox{ photons s}^{-1}{\rm cm}^{-2}{\rm sr}^{-1}\,.
    \label{F-QN}
\end{equation}
In the same sector, the Fermi-LAT telescope registered the following photon flux \cite{FermiLAT1}:
\begin{equation}
    \label{F-LAT}
    F_{\rm Fermi} = 10^{-4}\mbox{ photons s}^{-1}{\rm cm}^{-2}{\rm sr}^{-1}\,.
\end{equation}
By comparing this flux with our estimate (\ref{F-QN}) we conclude that the observed value by the Fermi telescope (\ref{F-LAT}) may be fully explained within the QN model of dark matter if the baryonic charge number obeys the constraint
\begin{equation}
    B \lesssim 10^{27}\,.
\end{equation}
This bound is consistent with the constraints from below (\ref{Bconstraint}) and above (\ref{UpperConstraint}). Thus, {\it QNs may produce a significant contribution to the galaxy $\gamma$ photon background}.

\subsubsection{Transition radiation}

The charged $\pi^\pm$ mesons produced in the hydrogen annihilation in the antiquark core may have energies in the range $100-400$ MeV. The corresponding relativistic factor is $\gamma_\pi\approx 1-3$. For a relativistic particle in the medium with plasma frequency $\omega_p$ the energy per unit frequency of transition radiation is \cite{Electrodynamics}
\begin{equation}
\label{trans-energy}
    \frac{d{\cal E}}{d\omega} = \frac{e^2}{\pi c}
    \left[
        \left(  1+2\frac{\omega^2}{\gamma_\pi^2 \omega_p^2} \right)
        \ln\left( 
        1+\frac{\gamma_\pi^2 \omega_p^2 }{\omega^2}\right) -2
    \right].
\end{equation}
This formula applies for frequencies $\omega\gtrsim 0.01 \gamma_\pi\omega_p$ \cite{Electrodynamics}, as at smaller frequencies it diverges. Consider $\omega =1$ MeV as a typical frequency of transition radiation in the case of quark nuggets with plasma frequency (\ref{omegap}). For this frequency, the energy per unit frequency (\ref{trans-energy}) is
\begin{equation}
    \frac{d{\cal E}}{d\omega} \approx 1.8 \frac{e^2}{\pi c}\,.
\end{equation}
Multiplying this energy by the $\pi$ meson emission rate in the bulge of our galaxy (\ref{F-rate}) we find the flux of transition radiation in the bulge of our galaxy produced by quark nuggets:
\begin{equation}
    F \frac{d{\cal E}}{d\omega} = 1.2\times 10^{47} B^{-1/3} {\rm s}^{-1}\,.
\end{equation}
The corresponding radiation spectral density observed at the distance 8.5 kpc from the galactic center would be
\begin{equation}
    P_{\rm QN} = \frac{F}{\Theta S} \frac{d{\cal E}}{d\omega}
    \approx \frac{1.4\times 10^{-26}}{B^{1/3}}
    \frac{\rm erg}{\mbox{s cm}^2\,\rm Hz\ sr}\,,
\label{PQN}
\end{equation}
where $\Theta$ and $S$ are given by Eqs.~(\ref{ThetaS})

It is instructive to compare the photon flux (\ref{PQN}) with the sensitivity of COMPTEL gamma-ray telescope \cite{COMPTEL}. At the frequency $\omega = 1$ MeV considered above, this detector has sensitivity of order
\begin{equation}
\label{PCOMPTEL}
    P_{\rm COMPTEL} \simeq 5\times 10^{-32} \frac{\rm erg}{\mbox{s cm}^2\,\rm Hz \ sr}\,.
\end{equation}
Comparing Eqs.~(\ref{PQN}) and (\ref{PCOMPTEL}) we conclude that the transition radiation from QNs could be detected if
\begin{equation}
    B\lesssim 5\times 10^{18}\,.
\end{equation}
Unfortunately, this constraint is incompatible with the allowed region for QN baryon number (\ref{Bconstraint}). Therefore, we conclude that {\it the transition radiation from quark nuggets in the bulge of our galaxy cannot be detected} by detectors like COMPTEL.


\section{Summary and discussion}
\label{Summary}

In this paper, we studied the thermal radiation and its manifestation in the quark nugget model of dark matter. Recall that the quark nuggets are compact composite objects composed of a (anti)quark core and an electron or positron cloud which provides electric neutrality of the nugget. The thermal radiation of quark nuggets originates from thermal fluctuations of density in this electron/positron cloud.

To determine the power of thermal radiation we represent quark nuggets as small spherical particles with a given dielectric constant. This dielectric constant determines reflection, refraction and absorption of electromagnetic radiation by such particles. The thermal emissivity of such particles is expressed via the radiation absorption cross section which is, in turn, a function of the size of the particle and the dielectric constant \cite{BHbook}. Within the Drude theory the dielectric constant is expressed in terms of the plasma frequency $\omega_p$ and damping constant $\gamma$ of the electron gas. In Sect.~\ref{DielectricConstant} we estimate these parameters in the case of the quark nugget model as $\omega_p\approx 2$ MeV, $\gamma\approx 475$ eV. Note that the QN effective temperature, which determines the thermal radiation spectrum, has a very weak sensitivity to these parameters ($T\propto \omega_p^{2/9} \gamma^{-1/9}$), so we do not need high accuracy in their calculations.

Having fixed the value of the dielectric constant, we calculate the thermal emissivity $E(\omega)$ of quark nuggets within the Mie theory, see, e.g., \cite{BHbook}. The thermal emissivity as a function of frequency is shown in Fig.~\ref{fig-emissivity} in different frequency bands. These plots show that $E(\omega)$ varies slowly within the interval $10^{-4}-10^{-3}$. It is also interesting to note that this function exhibits oscillations at particle's eigenfrequencies $\omega R/c=1,2,3,\ldots$ These oscillations are apparent at low frequencies on the plots of total radiation power of a quark nugget in Fig.~\ref{RadiationPower}. However, they may be unobservable if the quark nuggets have size and temperature distributions.

In Fig.~\ref{R-plots}, we compare the QN radiation intensity with the black body radiation. The ratio between these intensities is described by the function $R(T)$ which varies in the range $10^{-4}-10^{-3}$. In particular, at $T=1$ eV we find $R\approx 10^{-4}$. Thus, at this temperature the thermal radiation power from QNs is nearly 100 times higher than that estimated in Ref.~\cite{WMAPhaze}. This is the result of collective effects which have not been taken into account in Ref.~\cite{WMAPhaze}. 
The dependence of the radiation power $P(\omega,T)$ on the frequency $\omega$ is also very strongly affected by the collective effects. In Ref.~\cite{WMAPhaze} the spectrum is nearly ``flat'' in some range of frequencies while $P(\omega,T)=\pi E(\omega)I_0(\omega,T)$ has a steep decrease towards small frequencies as a combined effect of the decrease of the emissivity $E(\omega)$ and Plank  function $I_0(\omega,T)$. This is important for the explanation of such low-frequency phenomena as ``WMAP haze''  \cite{WMAP1,WMAP2,WMAP3,WMAP4}.

We revisit some astrophysical implications and predictions of the quark nugget model of dark matter.  We found that the effective temperature of quark nuggets in the interstellar medium in our galaxy should be of order $T\approx 0.2$ eV.
The thermal radiation spectrum has maximum of intensity at $\omega\approx 0.7$ eV, corresponding to infrared (IR) radiation.  
 We  considered thermal radiation in the IR spectrum and showed that its flux is significantly lower than the observed IR radiation in the galaxy by COBE telescope.
 
We reconsider the proposal \cite{WMAPhaze} that the quark nuggets may be responsible for the excess of radiation in the 22-93 GHz range from the bulge of our galaxy known as the ``WMAP haze'' \cite{WMAP1,WMAP2,WMAP3,WMAP4,WMAP5}. Within the QN model, there may be a few sources for this radiation: thermal radiation from density fluctuations in the positron cloud, bremsstrahlung radiation produced by energetic charged $\pi$ mesons passing through the positron cloud of QNs and synchrotron radiation from emitted electrons and positrons which pass through random magnetic fields in the galaxy (annihilation of hydrogen on QN produces charged pions which decay to electrons and positrons via intermediate muons). We estimated the flux of all these types of radiation and came to the  conclusion that its value is significantly lower than that observed by WMAP.

We stress that the negative results regarding possibilities to detect the thermal radiation from quark nuggets in our galaxy do not allow us to rule out these possibilities.
Indeed, the spectrum and intensity of thermal radiation from antiquark nuggets strongly depend on QN's effective temperature. This temperature is determined by the annihilation rate of antiquark nuggets with visible matter which is represented mainly by the hydrogen gas in the interstellar medium. In this paper, we assume a smooth and monotonic distribution of the hydrogen gas in the galaxy (\ref{gas-density}). In reality, however, the gas density in the galaxy may be very non-uniform, as this gas is often clumped in clouds. In such clouds a higher flux of hydrogen falls to QN surface increasing annihilation rate and temperature of QN. According to our estimate, inside a gas cloud the temperature is 0.5 eV while outside it is 0.2  eV.  Consequently, the thermal radiation spectrum will be shifted towards higher energies. The maximum of the radiation intensity is in visible light spectrum and QN radiation may be observed by optical telescopes. It would be interesting to study the radiation in the visible spectrum from dense molecular clouds in different areas of our galaxy. The intensity of this radiation should correlate with dark matter density distribution. 

If the density of clumps of dark matter particles is correlated with the distribution of interstellar gas clouds, the combined quark nuggets annihilation energy may be significantly higher. This increases both thermal and  non-thermal radiation of QN. We leave this problem for further studies.

Although the thermal radiation from QNs in the galaxy seems to be too low to be detected, we show that a strong radiation from quark nuggets in the interstellar medium is produced by $\pi^0$ mesons which decay into $\sim200$ GeV photons. We estimated the flux of such photons produced by QNs and compared it with the observed $\gamma$ ray background radiation in the central region of our galaxy \cite{FermiLAT1}. We demonstrated that for $B\lesssim 10^{27}$ QNs may produce a significant contribution to the galaxy $\gamma$ photon background radiation observed by the Fermi telescope.

Another type of potentially observed radiation from QNs appears in MHz range of frequencies. This radiation has a synchrotron origin, as it is produced by ultrarelativistic electrons or positrons  propagating through random magnetic fields in the galaxy. We show that the flux of this radiation may by comparable with the one measured by the RAE1 satellite in our galaxy \cite{RF1}. Thus, it is plausible that a significant portion of the galactic RF background radiation may be attributed to annihilation of hydrogen on antiquark nuggets with ultrarelativistic electrons and positrons as the final products. 

Finally, in this paper we investigated also other types of radiation which could be produced by QNs in our galaxy and compared them with the corresponding observations. We considered  the transition radiation which appears when charged $\pi$ meson cross the boundary of positron cloud and showed that it appears significantly lower than the background and, thus, unobservable.


\subsection*{Acknowledgements}
We are grateful to Ariel Zhitnitsky for useful discussions.
This work was supported by the Australian Research Council Grants No.\ DP190100974 and DP200100150 and the Gutenberg Fellowship.


%

\end{document}